\documentclass[11pt,a4paper]{article}
\usepackage{jcappub}

\usepackage{multirow}
\usepackage{natbib}
\usepackage{aas_macros}
\usepackage[latin1]{inputenc}
\usepackage{amsmath}
\usepackage{color}
\usepackage{color}
\usepackage{enumerate}
\usepackage{multirow}

\title{Cosmic shear calibration with forward modeling}
\author{Claudio Bruderer,}
\author{Andrina Nicola,}
\author{Adam Amara,}
\author{Alexandre Refregier,}
\author{J\"org Herbel,}
\author{and Tomasz Kacprzak}
\affiliation{Department of Physics, ETH Zurich, Wolfgang-Pauli-Strasse 27, 8093 Z\"urich, Switzerland}
\emailAdd{claudio.bruderer@phys.ethz.ch}

\abstract{Weak Gravitational Lensing is a powerful probe of the dark sector of the Universe. One of the main challenges for this technique is the treatment of systematics in the measurement of cosmic shear from galaxy shapes. In an earlier work, \cite{Refregier:2014aa} have proposed the Monte Carlo Control Loops (MCCL) to overcome these effects using a forward modeling approach. We focus here on one of the control loops in this method, the task of which is the calibration of the shear measurement. For this purpose, we first consider the requirements on the shear systematics for a given survey and propagate them to different systematics terms. We use two one-point statistics to calibrate the shear measurement and six further one-point statistics as diagnostics. We also propagate the systematics levels that we estimate from the one-point functions to the two-point functions for the different systematic error sources. This allows us to assess the consistency between the systematics levels measured in different ways. To test the method, we construct synthetic sky surveys with an area of 1,700 deg$^2$. With some simplifying assumptions, we are able to meet the requirements on the shear calibration for this survey configuration. Furthermore, we account for the total residual shear systematics in terms of the contributing sources. We discuss how this MCCL framework can be applied to current and future weak lensing surveys.}


\begin{document}
\maketitle

\section{Introduction}

Weak lensing surveys give us important measurements of the late-time Universe. With the substantial increase in the size and quality of lensing driven experiments, this technique will continue to grow in importance in experimental cosmology \citep[for reviews see e.g.][]{Bartelmann:2001aa,Refregier:2003aa,Hoekstra:2008aa,Kilbinger:2015aa}. However, making high precision measurements of weak lensing using wide-field galaxy surveys remains challenging. One of the main observational challenges is the accurate measurement of weak lensing shear from galaxy shapes. To reach the required precision for upcoming and future data, several potential sources of systematic errors need to be controlled in the data analysis steps. Biases can arise for numerous reasons \citep[for overviews see e.g.][]{Massey:2013aa,Hoekstra:2016aa}. These include, for instance, errors in estimating the local point spread function (PSF) and correcting for it \citep[e.g.][]{Paulin-Henriksson:2008aa}, the selection of the sample of lensing galaxies, and noise bias in the image \citep[e.g.][]{Melchior:2012aa,Refregier:2012aa}, which can all lead to systematic errors in the measurement of galaxy shapes. Furthermore, these biases need not be independent, but can display interactions difficult to model analytically \citep{Kacprzak:2014aa}.

Currently, several wide-field galaxy surveys are underway. The Dark Energy Survey\footnote{http://www.darkenergysurvey.org/} (DES) has published results on their science verification (DES SV) data covering about 140 deg$^2$ \citep[][]{Jarvis:2016aa,Becker:2016aa,Abbott:2016aa}. The Kilo-Degree Survey\footnote{http://kids.strw.leidenuniv.nl/} (KiDS) have reported results on $\sim$450 deg$^2$ \citep{Hildebrandt:2017aa}. Hyper Suprime-Cam\footnote{http://hsc.mtk.nao.ac.jp/ssp/} (HSC) has made a first public data release \citep{Aihara:2017aa} with gravitational lensing measurements being presented later in the year. While KiDS reports a tension in $S_8\equiv\sigma_8\sqrt{\Omega_m/0.3}$ with the Planck 2015 results \citep{Hildebrandt:2017aa}, where $\sigma_8$ is the normalization of the matter power spectrum on scales of $8 h^{-1}$ Mpc and $\Omega_m$ is the present matter density of the Universe, first DES constraints are consistent with either \citep{Abbott:2016aa}. Measurements on the next data releases are important for addressing this tension.

There are several methods for measuring shear. In some, the second- and higher-order moments of the galaxy light distribution are measured \citep[e.g.][]{Bruderer:2016aa,Hoekstra:2016aa}. Other methods rely on fitting parametric models to galaxy images \citep[e.g.][]{Miller:2007aa,Zuntz:2013aa,Sheldon:2014aa,Jarvis:2016aa}. Due to the aforementioned complex origin of systematic errors in shear measurement, most of these methods share a need to be calibrated. This is typically achieved using image simulations \citep[e.g. for a recent analysis][]{FenechConti:2016aa}, although alternative approaches exist \citep[e.g.][]{Huff:2017aa}.

Our approach for calibrating shape measurement methods, called the Monte Carlo Control Loops (MCCL), was proposed by \cite{Refregier:2014aa}. They developed a framework of control loops set up to, in the first control loop, calibrate fast image simulations to match in key distributions of the data to be analyzed. In the second control loop, the lensing measurement is calibrated and performed on the simulations. And in the third control loop, Monte Carlo methods are used to vary the image simulations within a volume of parameter space where the simulations and the data are consistent. This is to assert that the calibrated shape measurement prescription is robust to these changes. This framework therefore calibrates the shear measurement and assesses its robustness. An initial implementation of the MCCL was presented by \cite[][henceforth B16]{Bruderer:2016aa} and applied to the DES SV. The MCCL were shown to calibrate a given shape measurement prescription for the corresponding biases to be sub-dominant to the measurement of, in that case, lensing one-point functions. The MCCL framework was also applied to the measurement of the redshift distribution of cosmological galaxy samples \citep{Herbel:2017aa}.

Here, we update the intermediate target set in B16 and aim at measuring lensing two-point functions. For this purpose, we set our targets so as to be statistics-limited by a 1,700 deg$^2$ wide-field survey. In order to meet the more stringent requirements of this measurement, we focus here on describing the second control loop, the shape measurement pipeline and its calibration.

This paper is organized as follows. In section~\ref{sec:systematicsdiagnostics} we describe the requirements on the calibration of the shear measurement needs to satisfy. Furthermore, we introduce our model for the contributions of different terms to the total systematic error. In section~\ref{sec:syntheticsurveys} we present our synthetic imaging surveys on which we calibrate the shear measurement. We then describe our shear measurement pipeline in section~\ref{sec:shearmsrmtmethod}. The tests we perform in order to evaluate whether the calibrated shear measurement satisfies the requirements are shown in section~\ref{sec:implementation}. The results are then discussed in section~\ref{sec:results}. Finally, we conclude in section~\ref{sec:discussion}.

\section{Requirements}\label{sec:systematicsdiagnostics}

\subsection{Requirements targets}\label{sec:requirements}
In this work, we focus on quantifying the requirements targets using the conventions set out in \cite{Amara:2008aa}. In this framework we define systematics contributions as a variance like term,

\begin{equation}\label{eq:sigmasyssqdef}
  \sigma_{\rm sys}^2 = \int  \frac{\ell (\ell+1)}{2 \pi} \left| C_\ell^{\rm sys} \right| ~d \ln \ell,
\end{equation}
where $C_\ell^{\rm sys}$ is the spherical harmonic power spectrum of the shear systematics. Furthermore, \cite{Amara:2008aa} give a scaling relation that allows us to set the target level for $C_\ell^{\rm sys}$ depending on the properties of the lensing survey being considered. 
\begin{equation}\label{eq:sigma_sys}
  \sigma_{\rm sys}^2 < 3.3 \times 10^{-6} \left( \frac{A}{ 1700 ~{\rm deg}^2} \right)^{-0.5} \left( \frac{n_g}{4~{\rm arcmin}^{-2}} \right)^{-0.5}\left( \frac{z_m}{0.6} \right)^{-0.6} \left( \frac{\ell_{\rm max}}{1000} \right)^{-.4},
\end{equation}
where $A$ is the area of the survey, $n_g$ the surface density of galaxies, and $z_m$ the median redshift of the galaxies. 
These requirements are designed to ensure that the systematic errors are sub-dominant compared to statistical errors for cosmological parameter constraints. The central value corresponds to the survey we consider in this paper, which has a statistical precision comparable to current generation weak lensing surveys. Therefore, this leads to a requirements target of $\sigma_{\rm sys}^2 \lesssim 3\times 10^{-6}$.

\subsection{Systematics terms}
First, we describe how the level of systematics can be estimated in order to check whether we meet the
requirements described in section~\ref{sec:requirements}. For this purpose, let us consider an estimator $\hat{\gamma}_i$ for the shear $\gamma_i$ (with $i=1,2$) which is related to the true shear by
\begin{equation}\label{eq:shearbias}
  \hat{\gamma}_i = (1+m)\gamma_i +c_i,
\end{equation}
where $m$ and $c_i$ quantify multiplicative and additive systematics respectively, and both can be spatially varying.

The systematics can arise from spatial variations for various sources, as for example: PSF ellipticity $e_{p,i}$, PSF size $r_p$, galaxy size $r_g$, galaxy magnitude $m_g$, galaxy S\`ersic index $n_g$, root mean square (rms) of the local background noise in the image $\sigma_{bg}$, and number of galaxies in close vicinity $N_g$. Note that these systematics sources can be scalar ($r_p$, $r_g$, $m_g$, $n_g$, $\sigma$, $N_g$) or spin-2 quantities ($e_{p,i}$). Let us collectively denote this set of quantities $\theta$ as
\begin{equation}
  {\Theta} = \{ e_{p,i},r_p,r_g,m_g,n_g,\sigma_{bg},N_g\}
\end{equation}
As a result, the multiplicative and additive contributions $m$ and $c_i$ can be functions of $\theta \in \Theta$, which are in turn functions of position ($x, y$).

In this work, we use the true PSF information at the galaxies' positions. The main goal of this paper is the description of our framework to assess calibration biases for MCCL and its application to a realistic test scenario, where an input shear power spectrum is recovered from noisy images. The PSF reconstruction in our framework will however be addressed in future work. The induced biases due to an imperfect PSF reconstruction in recovering the power spectrum have been studied extensively by e.g. \cite{Paulin-Henriksson:2008aa,Massey:2013aa}.

For a given systematics source $\theta \in \Theta$, the estimated 2-point component-wise shear correlation function is then
\begin{equation}
  \langle \hat{\gamma_i} \hat{\gamma_j} \rangle = \langle \gamma_i \gamma_j \rangle  + 2 \langle m(\theta) \rangle \langle \gamma_i \gamma_j \rangle + \langle m(\theta)m(\theta) \rangle \langle \gamma_i \gamma_j \rangle + \langle c_i(\theta) c_j(\theta)\rangle,
\end{equation}
so that the scalar 2-point shear correlation function is
\begin{equation}
  \langle \hat{\gamma} \hat{\gamma} \rangle = \sum_{i=1,2} \langle \hat{\gamma_i} \hat{\gamma_i} \rangle = \langle \gamma \gamma \rangle  + 2 \langle m(\theta) \rangle \langle \gamma \gamma  \rangle + \langle m(\theta)m(\theta) \rangle \langle \gamma \gamma \rangle + \langle c(\theta) c(\theta)\rangle
\end{equation}

The systematics variance can be estimated by
\begin{equation}
  \sigma_{\rm sys}^2=\langle \hat{\gamma}  \hat{\gamma} \rangle - \langle \gamma \gamma \rangle 
\end{equation}
at zero lag.

\subsection{Systematics from scalar sources}

Let us begin by considering scalar systematics sources that do not directly produce an additive term. We assume first that the different systematics sources are independent of each other. For such scalar sources $\theta$, the systematics variance would thus be given by
\begin{equation}\label{eq:sigmasyscalar}
  \sigma_{\rm sys}^2 = 2 \langle m(\theta) \rangle \langle \gamma \gamma \rangle + \langle m(\theta)m(\theta) \rangle \langle \gamma \gamma \rangle.
\end{equation}

\subsection{Systematics from Spin-2 sources}
 
For a spin-2 systematics source $\theta_i$, such as the PSF ellipticity $e_{p,i}$, let us consider only their additive contribution. We model this contribution to linear order as
\begin{equation}
  c_i(\theta) = \alpha \theta_i
\end{equation} 
where the leakage factor $\alpha$ is defined as
\begin{equation}
  \alpha \delta_{ij} = \frac{\partial \hat{\theta}_i}{\partial \theta_j}.
\end{equation}
We have assumed that $\alpha$ is a scalar or, in other words, that off-diagonal element in the leakage factor can be neglected. To go beyond this, $\alpha$ would need to be described by a tensor.

In the case we consider, the systematics variance is given by
\begin{equation}\label{eq:sigmasysvector}
  \sigma_{\rm sys}^2 =  \langle c(\theta) c(\theta)\rangle = \alpha^2  \langle \theta \theta \rangle.
\end{equation}

\section{Simulations}\label{sec:syntheticsurveys}

To test our method, we generate synthetic data from a known underlying model. A shear power spectrum is input to the image simulations. This data is then analyzed in order to recover the shear power spectrum and assess biases with the framework developed in section~\ref{sec:systematicsdiagnostics}.

We require four key ingredients in order to generate synthetic survey data: the cosmological model (section~\ref{sec:cosmology}), the astronomical models (section~\ref{sec:astronomy}), the survey strategy (section~\ref{sec:surveystrategy}), and modeling of instrumental effects (section~\ref{sec:obsinsteffects}). These are discussed in the stated sections below. The key steps in simulating an image then are outlined in section~\ref{sec:datageneration}.

\subsection{Cosmological model}\label{sec:cosmology}

We set our baseline parameters of the $\Lambda$CDM cosmological model to \{$\Omega_b h^2=0.022$, $\Omega_c h^2=0.12$, $100\theta_{MC}=1.0$, $\tau=0.066$, $\ln\left(10^{10}A_s\right)=3.1$ and $n_s=0.97$\}, so that they are close to the latest Planck constraints \cite{Planck-Collaboration:2015aa,Planck-Collaboration:2015ab}. The derived parameters are \{$h=0.68$, $\Omega_\Lambda=0.69$, $\Omega_m=0.31$, $\sigma_8=0.81$\}. To calculate the expected lensing signal in our mocks we also need to set the redshift distribution, $n(z)$, of the galaxies. We use the parametric form \citep[e.g.][]{Amara:2008aa,Smail:1994aa}
\begin{equation}\label{eq:nofz}
  n(z) = z^{a}\cdot\mathrm{Exp}[-(z/z_0)^{b}],
\end{equation}
where $a$, $b$, and $z_0$ are free parameters. We set the parameters of the distribution to $a=2.2$, $b=1.7$, and $z_0=0.42$ to match expectation based on the DES SVA1\footnote{http://des.ncsa.illinois.edu/releases/sva1D} lensing sample \citep{Jarvis:2016aa,Bonnett:2016aa}.

We calculate the E-mode theory power spectrum for the shear signal using \textsc{PyCosmo} \citep[][Refregier et al., in prep]{Refregier:2011aa} and set the B-mode power spectrum to zero. We then generate Gaussian random field realizations of \textsc{HEALPix} \citep{Gorski:2005aa} maps with $\textsc{nside} = 1024$, which are additionally convolved with the \textsc{HEALPix}-pixel window function.

\subsection{Astronomical models}\label{sec:astronomy}

\begin{table}[tbp]
  \begin{tabular}{c|l|l|l}
     & Parameter & Value & Description \\

    \hline

    \multirow{11}{*}{\rotatebox[origin=c]{90}{Galaxy model}} & \multirow{2}{*}{$a_0$, $a_1$, $a_2$} & \multirow{2}{*}{4.52, 0.356, -0.0076} & Coefficients of the cumulative magnitude \rule{0pt}{2.5ex} \\
     & & & distribution (equation~\ref{eq:magcumdist}) \\
     & \multirow{2}{*}{$\theta_g$} & \multirow{2}{*}{0.14} & Rotation angle correlating \rule{0pt}{2.5ex} \\
     & & & magnitude and size \\
     & $\sigma_g$ & 0.24 & Width of the log-normal size distribution \rule{0pt}{2.5ex} \\
     & $mag_{piv}$, $r_{piv}$ & 25.309, 0.160 arcsec & Pivot point in size-magnitude plane \rule{0pt}{2.5ex} \\
     & \multirow{2}{*}{$\mu_1$, $\sigma_1$, $\mu_2$, $\sigma_2$, $d_1$} & \multirow{2}{*}{0.3, 0.5, 1.6, 0.4, 0.2} & S\`ersic index distribution for bright \rule{0pt}{2.5ex} \\
     & & & galaxies ($m_g\leq20$; equation~\ref{eq:sersicdist}) \\
     & \multirow{2}{*}{$\mu_3$, $\sigma_3$, $d_3$} & \multirow{2}{*}{0.1, 1, 0.2} & S\`ersic index distribution for faint \rule{0pt}{2.5ex} \\
     & & & galaxies ($m_g>20$; equation~\ref{eq:sersicdist}) \\
     & $\sigma_e$ & 0.39 & Rms of the Gaussian distributions for $e_i$ \rule{0pt}{2.5ex} \\

    \hline

    \multirow{9}{*}{\rotatebox[origin=c]{90}{Survey design}} & $A_{survey}$ & 1,700 deg$^2$ & Survey area \rule{0pt}{2.5ex} \\
     & $N_{images}$ & 3,296 & Number of images \rule{0pt}{2.5ex} \\
     & $N_{pix}$ & 10,000$\times$10,000 & Number of pixels in each image \rule{0pt}{2.5ex} \\
     & pixel scale & 0.263'' & Size of the pixels in each image \rule{0pt}{2.5ex} \\
     & $mag_{0}$ & 30.2 & Target magnitude zero point \rule{0pt}{2.5ex} \\
     & \multirow{2}{*}{$t_{exp}$} & \multirow{2}{*}{90 s} & Exposure time of equivalent \rule{0pt}{2.5ex} \\
     & & & single exposure images \\
     & \multirow{2}{*}{$n_{exp}$} & \multirow{2}{*}{4} & Equivalent number of single exposures \rule{0pt}{2.5ex} \\
     & & & per coadded image \\

    \hline

    \multirow{13}{*}{\rotatebox[origin=c]{90}{Systematics model}} & \multirow{3}{*}{$\bar{d}_{FWHM}$, $\bar{e}_1$, $\bar{e}_2$} & \multirow{3}{*}{1.03'', 0.01, 0.015} & Approx. average values for the PSF \rule{0pt}{2.5ex} \\
     & & & FWHM and ellipticity estimated on \\
     & & & DES SVA1 data \\
     & $A_d$, $\ell_{0,d}$, $\ell_{1,d}$, $a_d$ & $4\cdot10^{-6}$, 35, 150, 0.8 & \multirow{2}{*}{$C_{\ell,d_{FWHM}}$ model parameters} \rule{0pt}{2.5ex} \\
     & $b_d$, $c_d$, $d_d$, $e_d$ & 4, -1.4, 2.5, 1.2 & \\
     & $A_e$, $\ell_{0,e}$, $\ell_{1,e}$, $a_e$ & $4\cdot10^{-8}$, 25, 150, 1.2 & \multirow{2}{*}{$C_{\ell,E}$ and $C_{\ell,B}$ model parameters} \rule{0pt}{2.5ex} \\
     & $b_e$, $c_e$, $d_e$, $e_e$ & 3, -1.4, 2.0, 0.5 & \\
     & $\beta$ & 2.5 & $\beta$ parameter of the PSF Moffat profile \rule{0pt}{2.5ex} \\
     & Gain & 4.4 $e^{-}$/ADU & Gain \rule{0pt}{2.5ex} \\
     & \multirow{3}{*}{$\sigma_{bg}$} & \multirow{3}{*}{$0.660 \pm 0.018$} & Mean and standard deviation of the \rule{0pt}{2.5ex} \\
     & & & normal distribution the rms values of the \\
     & & & Gaussian background noise are drawn from \\
     & Saturation & 45,000 ADU & Saturation value \rule{0pt}{2.5ex} \\

  \end{tabular}
  \caption{\label{tab:syntheticsurveytable} Baseline galaxy model, survey design, and systematics prescriptions. The parameter values and descriptions are summarized, while further details can be found in the text and in B16.}
\end{table}

We follow the prescriptions described in \cite{Berge:2013aa} and B16 to generate synthetic catalogs containing the intrinsic properties of the simulated galaxies and stars. For a given area $A$, we calculate the expected number of objects using the relation
\begin{equation}\label{eq:magcumdist}
  \log_{10}(N<m_g)=\sum_{i=0}^{2} a_i(m_g - 23)^i,
\end{equation}
where the coefficients $a_i$ are given in table~\ref{tab:syntheticsurveytable}. We set the magnitude limit to $m_g\leq27$, well below the detection threshold, since faint, unresolved objects can affect the calibration of the shear measurement \citep{Hoekstra:2015aa}. The number of galaxies actually assigned to a given area, $N$, is selected by drawing from a Poisson distribution centered on the expected value. The galaxies are then assigned intrinsic properties: position (RA, DEC); apparent magnitude $m_g$; half-light radius $r_g$; S\`ersic index $n_g$; ellipticity ($e_1$, $e_2$); and shear ($\gamma_1$, $\gamma_2$).

The positions are sampled uniformly. The shears are assigned based on the maps from section \ref{sec:cosmology}. The sizes and magnitudes of galaxies are strongly correlated and we follow the prescriptions in \cite{Berge:2013aa} and B16 for jointly drawing these two variables. The parameters for this step, \{ $\theta_g, ~mag_{piv}, ~\mathrm{log}_{10}r_{piv}$\}, are given in table \ref{tab:syntheticsurveytable}. The galaxy intrinsic ellipticity components $e_i$ are drawn independently from a Gaussian distribution $\mathcal{N}(0, \sigma_e)$, where values of $e_1^2+e_2^2$ exceeding 1 are rejected. In this work, we employ the ellipticity definition with $e=(a^2-b^2)/(a^2+b^2)$, where $a$ and $b$ are an ellipses' semi-major and semi-minor axes \citep[see e.g.][]{Rhodes:2000aa}.

Galaxies are modeled with a S\`ersic profile \citep{Sersic:1963aa} described by a single parameter $n_g$. We sample the following distributions for the S\`ersic indices depending on the galaxy's magnitude \citep{Berge:2013aa}
\begin{equation}\label{eq:sersicdist}
  f(n) = 
  \begin{cases}
  \exp\left[\mathcal{N}(\mu_1, \sigma_1)+\mathcal{N}(\mu_2, \sigma_2)\right] + d_1, & \text{for } m_g\leq 20 \\
  \exp\left[\mathcal{N}(\mu_3, \sigma_3)\right] + d_3, & \text{for } m_g > 20
  \end{cases},
\end{equation}
where $\mathcal{N}(\cdot, \cdot)$ is the normal distribution, $\mu_i$ and $\sigma_i$ the corresponding mean and standard deviation, and $d_i$ the offset values. The values are given in table~\ref{tab:syntheticsurveytable}.

We model stars in the images using a stellar population synthesis of the Galaxy \citep{Robin:2003aa} to generate apparent magnitude distributions in the $r$-band for different locations of the Sky. While the number of stars varies spatially across the survey area, the stars are placed within the individual images with a uniform probability.

\subsection{Survey strategy}\label{sec:surveystrategy}

We simulate a synthetic survey with a total area of about 1,700 deg$^2$. We divide the region into 3,296 individual images, where the images overlap by $\sim$100 pixels. The image dimensions are 10,000$\times$10,000 pixels, with a pixel scale of 0.263 arcsec\footnote{http://www.ctio.noao.edu/noao/node/2250}. We directly simulate stacked images with an equivalent number $n_{exp}=4$ of single exposures \citep[for details see][]{Berge:2013aa}. The specific key properties of this simulated survey are given in table~\ref{tab:syntheticsurveytable}.

\subsection{Observational and instrumental effects}\label{sec:obsinsteffects}

We model the point spread function (PSF) using an elliptical Moffat distribution described by four parameters \citep{Moffat:1969aa}. These are the half-light radius $r_{50,p}$, $\beta$, which describes the shape of the profile, and the two ellipticity components $e_{1,p}$ and $e_{2,p}$. Size and ellipticity are also allowed to vary across the survey area, for which we build a model based on the DES SVA1 data \citep{Jarvis:2016aa}. We use \textsc{PolSpice}\footnote{http://www2.iap.fr/users/hivon/software/PolSpice/} \citep{Szapudi:2001aa,Chon:2004aa} to first measure the three power spectra: $C_{\ell,d_{FWHM}}$, which is the angular power spectrum of PSF size variations, and $C_{\ell,E}$ and $C_{\ell,B}$, which are the E- and B-modes of the PSF ellipticity. We model these power spectra using the form, 
\begin{equation}\label{eq:psfmodel}
  C_{\ell} = A\ell^{a} \left[1 + \left(\frac{\ell}{\ell_0}\right)^b \right]^c\left[1 + \left(\frac{\ell}{\ell_1}\right)^d \right]^e.
\end{equation}
The parameters for the models employed in this work are given in table~\ref{tab:syntheticsurveytable}. For simplicity we set the E- and the B-mode power spectra within our model to be the same. We then generate Gaussian random field realizations of these angular power spectra leading to maps for $d_{FWHM}$, $e_1$, and $e_2$, where we set $\textsc{nside} = 1024$, and convolve the map with the pixel window function. The Gaussian random fields have a mean of zero for every quantity. Since on the DES SVA1 PSF maps we have estimated the average PSF ellipticities to be $e_1\sim0.01$ and $e_2\sim0.014$, we adjust the simulated PSF ellipticity maps by shifting them by this value. We set the PSF FWHM to 0.93 arcsec and the PSF Moffat parameter $\beta$ to 2.5, which also allows us to convert from $d_{FWHM}$- to a $r_{50,p}$-map. The PSF properties of a position within a \textsc{HEALPix} pixel are set to the value at its center. The resulting maps are shown in appendix~\ref{sec:creatingpsfmaps}.

Similar to B16, we model the background noise with a Gaussian noise component. Its rms values $\sigma_{bg}$ are spatially varying, and they are drawn from the normal distribution $\mathcal{N}(0.660, 0.018)$, describing typical values. We furthermore include a prescription for saturation \citep[for details see][]{Berge:2013aa}.

\subsection{Image generation}\label{sec:datageneration}

We simulate images using the Ultra Fast Image generator \citep[UFig;][]{Berge:2013aa}. As mentioned above, UFig directly generates coadded images bypassing the generation of single exposure images \citep[for details see][]{Berge:2013aa}. The key simulation steps for each of the $N_{images}$ images of our synthetic survey are: (i) generate an image containing only background noise, (ii) generate galaxy and star catalogs with the models described in section~\ref{sec:astronomy}, (iii) evaluate the PSF at the object positions (section~\ref{sec:obsinsteffects}), (iv) evaluate the shear at the galaxy positions, and (v) render the objects convolved with the local PSF profile and add them to the image containing the background noise.

In order for the background noise properties of the simulations to be comparable to proper coadded images, we add the Gaussian background noise (section~\ref{sec:obsinsteffects}) first, and then additionally resample the whole image with an empirically derived convolution kernel. This kernel mimics correlations between the pixel values due to the coaddition of $n_{exp}$ single exposures. This is similar to the Lanczos convolution used in B16 and \cite{Berge:2013aa}. Furthermore, the image contains photon noise. Galaxies are sampled photon-by-photon and thus naturally include it, while for the stars, which are simulated on a pixel grid, we add Poissonian noise. Further details on the image generation process can be found in \cite{Berge:2013aa}.

\section{Measurement method}\label{sec:shearmsrmtmethod}

This section describes the analysis steps used to measure the angular shear power spectrum on coadded images. The images are first analyzed with \textsc{SExtractor} \citep[][henceforth SE]{Bertin:1996aa}, where the values for the keys \textsc{GAIN}, \textsc{PSF\_FWHM}, and \textsc{SATURATE} are individually set to the coadded images' input values. Then, we perform a star-galaxy separation (section~\ref{sec:sgseparation}) and select the galaxy sample (section~\ref{sec:galaxysample}). On these selected galaxies, their shear is estimated with a calibrated prescription based the moments of the galaxies' light distribution (section~\ref{sec:shearmsrmt}). Finally, sky maps of the estimated shear are constructed and the angular power spectrum is estimated (section~\ref{sec:angularpowerspectrum}).

\subsection{Star-galaxy separation}\label{sec:sgseparation}

Stars and galaxies are separated in the \textsc{MAG\_AUTO} vs. \textsc{MU\_MAX} plane (e.g. \cite{Leauthaud:2007aa}) with linear relations, empirically calibrated on our simulations. Galaxies are selected with
\begin{equation}
  \begin{aligned}
    20.5 \leq \textsc{MU\_MAX} &: \ \ \textsc{MU\_MAX} \geq \mu \cdot \textsc{MAG\_AUTO} + \nu + 0.7, \\
    16.5 < \textsc{MU\_MAX} < 20.5 &: \ \ \textsc{MU\_MAX} > \mu \cdot \textsc{MAG\_AUTO} + \nu - 0.5,
  \end{aligned}
\end{equation}
where $\mu = 0.9892$ and $\nu = 0.5158$. Values of \textsc{MU\_MAX} below 16.5 are mainly saturated objects and hence excluded.

In this work, we describe a framework to evaluate the performance of the calibration of the shear measurement and predict residual systematics. Since we use intrinsic properties of galaxies as sources inducing systematics (see section~\ref{sec:implementation}), we remove objects such as stars that have been misidentified as galaxies for this paper. The effect of star contamination to the sample will be explored in future work.

\subsection{Galaxy sample selection}\label{sec:galaxysample}

In order to construct the lensing sample, we perform cuts on Signal-to-Noise (S/N), the size of the galaxies, and a cut on the size of galaxies relative to the local PSF size, similar to the cuts in B16. The first two cuts remove galaxies with
\begin{equation}
  \frac{\textsc{FLUX\_AUTO}}{\textsc{FLUXERR\_AUTO}} \leq 15,
\end{equation}
and which exceed an upper threshold in the half-light radius
\begin{equation}
  \textsc{FLUX\_RADIUS} \geq 30 \ \mathrm{Pixels}.
\end{equation}
For the latter, we convert the local values of the PSF size and ellipticity to normalized quadrupole moments $P_{ij}$, and then use the diagonal elements $P_{11}$ and $P_{22}$. We require the galaxies in the sample to satisfy
\begin{equation}
  \sqrt{\frac{\textsc{FLUX\_RADIUS}^{2}}{(P_{11}+P_{22})\cdot2\mathrm{ln}2}} \geq 1.2.
\end{equation}

\subsection{Shear measurement and PSF correction}\label{sec:shearmsrmt}

We follow the prescriptions outlined in B16, which are themselves to first order based on \cite{Rhodes:2000aa}, to estimate the galaxies' shear. Similar to B16, the measurement is complexified only if needed; that is, if the accuracy of the calibrated shear measurement does not satisfy the requirements targets set in section~\ref{sec:requirements}. The complexification can be achieved by e.g. including higher-order terms.

First, we convert the local PSF size and ellipticity \{$r_p$, $e_{p,i}$\} at each galaxy's position to quadrupole moments \{$P_{ij}$\}. These moments, scaled with a calibration factor $\lambda$, are then subtracted from the galaxy's weighted quadrupole moments as estimated by SE to perform an effective PSF-deconvolution
\begin{equation}
  \begin{aligned}
    J_{11} &= \textsc{X2WIN\_IMAGE} - \lambda P_{11}, \\
    J_{12} &= \textsc{XYWIN\_IMAGE} - \lambda P_{12}, \\
    J_{22} &= \textsc{Y2WIN\_IMAGE} - \lambda P_{22}.
  \end{aligned}
\end{equation}

The (uncalibrated) estimated shear is then
\begin{equation}
  \begin{aligned}
    \widetilde{\gamma}_{1} &= \left(\frac{J_{11}-J_{22}}{J_{11}+J_{22}}\right)\cdot\left(\frac{P_{11}+P_{22}}{J_{11}+J_{22}}\right)^{\beta-1}, \\
    \widetilde{\gamma}_{2} &= \left(\frac{2J_{12}}{J_{11}+J_{22}}\right)\cdot\left(\frac{P_{11}+P_{22}}{J_{11}+J_{22}}\right)^{\beta-1}.
  \end{aligned}
\end{equation}
Here, $\beta$, which creates an explicit dependence of the estimated shear on the ratio of galaxy and PSF size, is set to 1.3. We then require
\begin{equation}
  |\widetilde{\gamma}_{i}| < 1,
\end{equation}
and remove galaxies not satisfying this condition from the sample.

The SE windowed quadrupole moments \textsc{X2WIN\_IMAGE}, \textsc{Y2WIN\_IMAGE}, and \textsc{XYWIN\_IMAGE} are weighted with a circular Gaussian with rms $\textsc{FLUX\_RADIUS}/\sqrt{2\ln 2}$. This causes the shape estimate to be biased, which we calibrate with a single multiplicative factor $\eta$, i.e.
\begin{equation}
  \hat{\gamma}_{i} = \eta^{-1} \cdot \widetilde{\gamma}_{i}.
\end{equation}

The two calibration parameters $\lambda$ and $\eta$ are estimated jointly. The optimum values minimize the dependence of the estimated shear signal on the shape of the PSF (PSF leakage), which arises due to an insufficient PSF correction, and the discrepancy between the true and the estimated shear signal. These values are calibrated on a multiple of the target area. In our case, we use 10 surveys.

\subsection{Angular power spectrum estimation}\label{sec:angularpowerspectrum}

To estimate the angular power spectrum, we perform the following three steps. First, we create \textsc{HEALPix} maps of the estimated shear signal. We then use \textsc{PolSpice} to estimate the angular power spectra $C_{\ell,uncorr}$, which contain noise contributions due to the random alignment of the intrinsic shapes of galaxies (shape noise). Lastly, these noise contributions are removed.

We compute \textsc{HEALPix} maps of the estimated shear signal by putting each galaxy based on its angular coordinates into a pixel of a $\textsc{nside} = 1024$ map. We then average the shear values within each pixel. We note that since we simulate a synthetic survey consisting of overlapping images, objects at the edges of the images can be detected multiple times. In these cases, we randomly select one out of the ensemble of multiple detections, and remove the others from the sample.

To correct for the finite survey mask, \textsc{PolSpice} first estimates the pseudo-power spectrum on the masked maps. It then applies a Fourier transform to compute the real space correlation function corresponding to this power spectrum. This correlation function is then corrected for the limited survey area by dividing by the correlation function of the mask. \textsc{PolSpice} then applies a Fourier transform to it and returns the estimated power spectrum. The correlation function can only be estimated on angular scales smaller than the angular extent of the survey. We therefore set for the maximum separation angle of galaxy pairs considered in this computation $\textsc{thetamax} = 60 \ \mathrm{deg}$. The demasking procedure induces Fourier ringing artifacts, which can be reduced by applying a Gaussian apodization to the correlation function. We follow the prescriptions proposed by \cite{Chon:2004aa,Nicola:2016aa} and set the scale factor \textsc{apodizesigma} to $\textsc{thetamax} / 2$.

As a second parameter, we need to set \textsc{nlmax}, the maximum $\ell$ value for which the angular power spectra are computed. We choose $\textsc{nlmax}=$ 2,500. Furthermore, we correct for the effect of the \textsc{HEALPix} pixel window function.

To remove the contribution to $C_{\ell,uncorr}$ due to shape noise, we first estimate the noise level by shuffling the estimated shear values $\hat{\gamma}_{1}$ and $\hat{\gamma}_{2}$ among all galaxies in the sample. The positions are kept unchanged, and thus the patterns in the number density of the galaxies are not affected. Then, we perform the map-making and power spectrum estimation steps described above. In total, we repeat this process $M = 100$ times. These random realizations are then averaged, and finally subtracted from the estimated angular power spectrum
\begin{equation}
  \hat{C}_{\ell} = C_{\ell,uncorr} - \frac{1}{M}\sum_{i=1}^{M} C_{\ell,noise,i}.
\end{equation}

\section{Measuring the systematic errors}\label{sec:implementation}

In this section we present the implementation of the diagnostic tests introduced in section~\ref{sec:systematicsdiagnostics}. These check that the calibrated shear measurement (see section~\ref{sec:shearmsrmtmethod}) satisfies the requirements laid out in section~\ref{sec:requirements}.

To assess the accuracy of a measurement, one is limited by the statistical uncertainties of the measurement. To circumvent this, a number of realizations $N$ are required to reduce the statistical uncertainties. We empirically find that, similar to B16, to evaluate whether the shear measurement satisfies the requirements, the number needs to be $N \gtrsim 100$. Since we estimate the shear signal on a synthetic survey with an area of 1,700 deg$^2$ and place requirements on this measurement, we set $N=250$. This corresponds to a total of roughly 425,000 deg$^2$, equivalent to 10 full skies.

Furthermore, we choose to limit our analysis to a maximum multipole of $\ell_{\rm max}=$ 1,000 for several reasons. The contribution to the power spectrum due to the intrinsic shape is comparable to the signal on these small scales. This leads to the power spectrum estimation being more susceptible to the correction of the shape noise. Furthermore, the uncertainties in correcting for the pixel window function of the \textsc{HEALPix} maps increase with the multipole. For instance, the correction on the power spectrum for scales $\ell\sim$ 1,000 are at the level of about 5\%. Finally, the theoretical interpretation of the power spectrum is more challenging for larger multipoles $\ell$. Due to small angular scales probing the non-linear evolution of the density field, correlations between modes are increasingly significant for multipoles $\ell \gtrsim$ 1,000 \citep[e.g.][]{Harnois-Deraps:2012aa}

The detailed implementation of the systematics diagnostics is presented below. In section~\ref{sec:sigmasysshear}, we first describe how $\sigma_{\rm sys}^2$ is evaluated directly by measuring the difference between the true and the estimated shear angular power spectra. We then present the estimation for the residual systematics in sections~\ref{sec:sigmasyscatalog} \&~\ref{sec:sigmasyseval}. Our method relies on first separately mapping out the sensitivities of the shear measurement to the distribution of each systematics source. Based on these, we then predict the value of $\sigma_{\rm sys}^2$. This diagnostic tool is important, as, given the prediction agrees with the residual systematic errors, it allows us to prioritize which sensitivities need to be decreased in order to satisfy the requirements and minimize $\sigma_{\rm sys}^2$.

\subsection{Direct \texorpdfstring{$\sigma_{\rm sys}^2$}{sigma-sq} evaluation}\label{sec:sigmasysshear}

For each of the 250 simulated surveys, the shape noise-corrected shear power spectrum $\hat{C}_{\ell, j}$ is estimated (see section~\ref{sec:angularpowerspectrum}). We can then interpret the residual between the mean of these power spectra and the true shear signal $C_\ell^t$ as the contribution to the angular power spectrum due to systematics terms
\begin{equation}\label{eq:clrecovery}
  C_\ell^{\rm sys} = \frac{1}{N}\sum_{j=1}^{N}\hat{C}_{\ell, j} - C_\ell^t.
\end{equation}
The corresponding value of $\sigma_{\rm sys}^2$ is computed with equation~\ref{eq:sigmasyssqdef}.

We use \textsc{PolSpice} with the settings described in section~\ref{sec:angularpowerspectrum} to compute the shear signal $C_\ell^t$ of the true shear maps. On these maps, we apply the survey mask. By comparing the estimated to the power spectrum computed using the true shear maps, we avoid the need of including an uncertainty due to cosmic variance in the analysis. We furthermore reduce the effects of demasking by applying the same survey mask to the true shear maps.

\subsection{Creating systematics maps}\label{sec:sigmasyscatalog}

To estimate the level of systematics, we construct maps. These maps capture the impact of different systematics terms on the shear measurement. Since these are maps of the intrinsic quantities of the galaxies, we first begin by matching the detected to the input catalog. This is done by ensuring that the positions are within 2.5 image pixels and the magnitude difference is less than 4.

We construct systematics maps for the six scalar sources: $r_g$, $m_g$, $n_g$, $r_p$, $\sigma_{bg}$, and $N_g$. For each of these systematics sources we follow the same steps.

First, a binning scheme in the corresponding quantity is derived from one of the survey realizations. We choose to bin in 19 bins and assert that each bin contains approximately the same number of objects. Choosing too many bins increases the noise in estimating the systematics impact of each diagnostic, while too few lead to an underestimation of the impact. We empirically find 19 to be a good compromise. We then apply this fixed binning scheme to the other random realizations. Within each bin $k$ in this scalar quantity, we estimate the multiplicative biases $m_{i,k}$ for each component using a linear relation
\begin{equation}
  \hat{\gamma}_{i} = (1+m_{i,k})\gamma^{t}_{i},
\end{equation}
where $\gamma^{t}_{i}$ is the true value for the i-th shear component. We then check that the two components $m_{1,k}$ and $m_{2,k}$ yield similar values. By taking the mean, these are then combined to a single value for the multiplicative bias $m_{k}$ in each bin. Lastly, we assign this value for the multiplicative bias to each galaxy in this bin. This allows us then to create three systematics maps: $\Delta m$, the average value of the multiplicative biases $m_{k}$ for each galaxy in each pixel; $\Delta m\gamma^t$, the spin-2 maps of the average values for the combination of the each galaxy's multiplicative bias and true shear values $\gamma^t_i$; and $\Delta\theta/\bar{\theta}$, the map of the relative deviation of the scalar quantity $\theta$ relative to the full samples' average value.

Similarly, we construct diagnostic maps for the spin-2 systematics sources, here the value of the PSF ellipticity at the galaxy's position $e_{i,p}$. We again first derive a binning scheme for each of the components. We choose here 1,000 bins, which we empirically find to be a good choice, each containing the same number of objects ($\sim$ 20,000). For each bin we calculate the mean and rms deviation of the shear estimate to the true value $(\hat{\gamma}_{i}-\gamma^t_{i})$. We then estimate the slope $\alpha_i$ for each component, i.e.
\begin{equation}\label{eq:spin2shearfit}
  \hat{\gamma}_{i}-\gamma^t_{i} = \alpha_i\theta_i,
\end{equation}
where $\theta_i$ is the i-th component of the spin-2 quantity. We perform a $4\sigma$-clipping in each bin to remove outliers, and then refit. This gives more robust estimates of $\alpha_i$. The slopes $\alpha_{i}$ are then added in quadrature to obtain a single value $\alpha$ for both components. Finally, we construct the systematics maps: $\theta$, the average value of each component of the spin-2 quantity; and $\alpha\theta$, the average value of each component and the leakage $\alpha$.

\subsection{Evaluation of \texorpdfstring{$\sigma_{\rm sys,s}^2$}{sigma-sq} for systematics diagnostics}\label{sec:sigmasyseval}

To reduce the noise in the estimation of the impact of each systematics source (section~\ref{sec:systematicsdiagnostics}), we perform the analysis on a combination of the diagnostics maps computed for every survey realization.

For each systematics source, we first average the diagnostic maps of the $N$ realizations. We then smooth each of the average diagnostic maps by applying a Gaussian kernel with a rms of $\pi/\ell_{smooth}$, where this smoothing scale is chosen such that $\ell_{smooth}\sim 2\ell_{max}\sim$ 2,000. Smoothing the diagnostic maps and the survey mask consistently allows us to make the power spectrum estimation of these systematics terms more robust for scales $\ell<\ell_{max}$. Furthermore, the two methods for evaluating $\sigma_{\rm sys,s}^2$ (see below), where the subscript `s' denotes that the maps are smoothed, can be made consistent (see below). We finally compute the angular power spectra of these maps using the same settings for \textsc{PolSpice} as given in section~\ref{sec:angularpowerspectrum}.

These combined diagnostic maps and power spectra allow us to evaluate the contributions to $\sigma_{\rm sys,s}^2$ for each individual scalar and spin-2 systematics source. To test for robustness, we use two methods to estimate the contributions. The first method uses combinations of the diagnostic maps to compute $\sigma_{\rm sys,s}^2$ (1pt), while the other uses the corresponding power spectra (2pt). Since a Gaussian kernel has been applied to the diagnostics maps, fluctuations at small angular scales are suppressed. Therefore, we ensure that the ranges in angular scales probed by the 1pt- and 2pt-method agree.

The contributions to $\sigma_{\rm sys,s}^2$ for scalar systematics sources (see equation~\ref{eq:sigmasyscalar}) can be divided into two terms $\sigma_{\rm sys,s,1}^2$ and $\sigma_{\rm sys,2}^2$
\begin{equation}\label{eq:sigmasyscalarimplementationfirst}
  \sigma_{\rm sys,s}^2 = \underbrace{2\langle m(\theta)\rangle\langle\gamma\gamma\rangle}_{\sigma_{\rm sys,s,1}^2} + \underbrace{\langle m(\theta)m(\theta)\rangle\langle\gamma\gamma\rangle}_{\sigma_{\rm sys,s,2}^2}.
\end{equation}

For the first method (1pt), we use the combined and smoothed diagnostic maps $\Delta m$ and the input shear maps of each component $\gamma_{i}^{t}$. We evaluate the first term $\sigma_{\rm sys,s,1}^2$ by summing over the non-empty pixels $N_{pix}$ and computing
\begin{equation}
  \sigma_{\rm sys,s,1}^2 = 2\left(\frac{1}{N_{pix}}\sum_j^{N_{pix}}\Delta m_j\right)\left(\frac{1}{N_{pix}}\sum_j^{N_{pix}}\left((\gamma_{1,j}^{t})^{2}+(\gamma_{2,j}^{t})^{2}\right)\right).
\end{equation}
And similarly for the second term
\begin{equation}
  \sigma_{\rm sys,s,2}^2 = \left(\frac{1}{N_{pix}}\sum_j^{N_{pix}}(\Delta m_j)^{2}\right)\left(\frac{1}{N_{pix}}\sum_j^{N_{pix}}\left((\gamma_{1,j}^{t})^{2}+(\gamma_{2,j}^{t})^{2}\right)\right).
\end{equation}

We estimate the effect of the scalar systematics for the second method (2pt), by computing the corresponding angular power spectra. These are then related using equation~\ref{eq:sigmasyssqdef} to $\sigma_{\rm sys,s}^2$. To estimate the first term $\sigma_{\rm sys,s,1}^2$, we compute the cross-correlation between the smoothed $\Delta m$ diagnostic maps and the true shear map
\begin{equation}
  C_{\ell}^{\rm sys} = 2\langle(\Delta m\gamma)\gamma\rangle.
\end{equation}
Similarly, we consider the auto-correlation of the $\Delta m$ diagnostic maps for the second term $\sigma_{\rm sys,2}^2$,
\begin{equation}\label{eq:sigmasyscalarimplementationlast}
  C_{\ell}^{\rm sys} = \langle(\Delta m\gamma)^2\rangle.
\end{equation}

For spin-2 systematics sources, the contributions to $\sigma_{\rm sys,s}^2$, which correspond to the auto-correlation of the additive bias $c$ (see equation~\ref{eq:sigmasysvector}), are estimated similarly for the two methods. For the first method again (1pt), we use the diagnostic maps for each component $\theta_{i}$ and compute
\begin{equation}\label{eq:sigmasysvectorimplementationfirst}
  \sigma_{\rm sys,s}^2 = \frac{1}{N_{pix}}\sum_j^{N_{pix}}\left((\alpha\theta_{1,j})^{2}+(\alpha\theta_{2,j})^{2}\right).
\end{equation}

While for the second method (2pt), we compute the auto-correlation of these diagnostics maps and use equation~\ref{eq:sigmasyssqdef}, i.e.
\begin{equation}\label{eq:sigmasysvectorimplementationlast}
  C_{\ell}^{\rm sys} = \langle(\alpha\theta)^2\rangle.
\end{equation}

\section{Results}\label{sec:results}
\begin{figure}[ht]
  \centering
  \includegraphics[width=.7\textwidth]{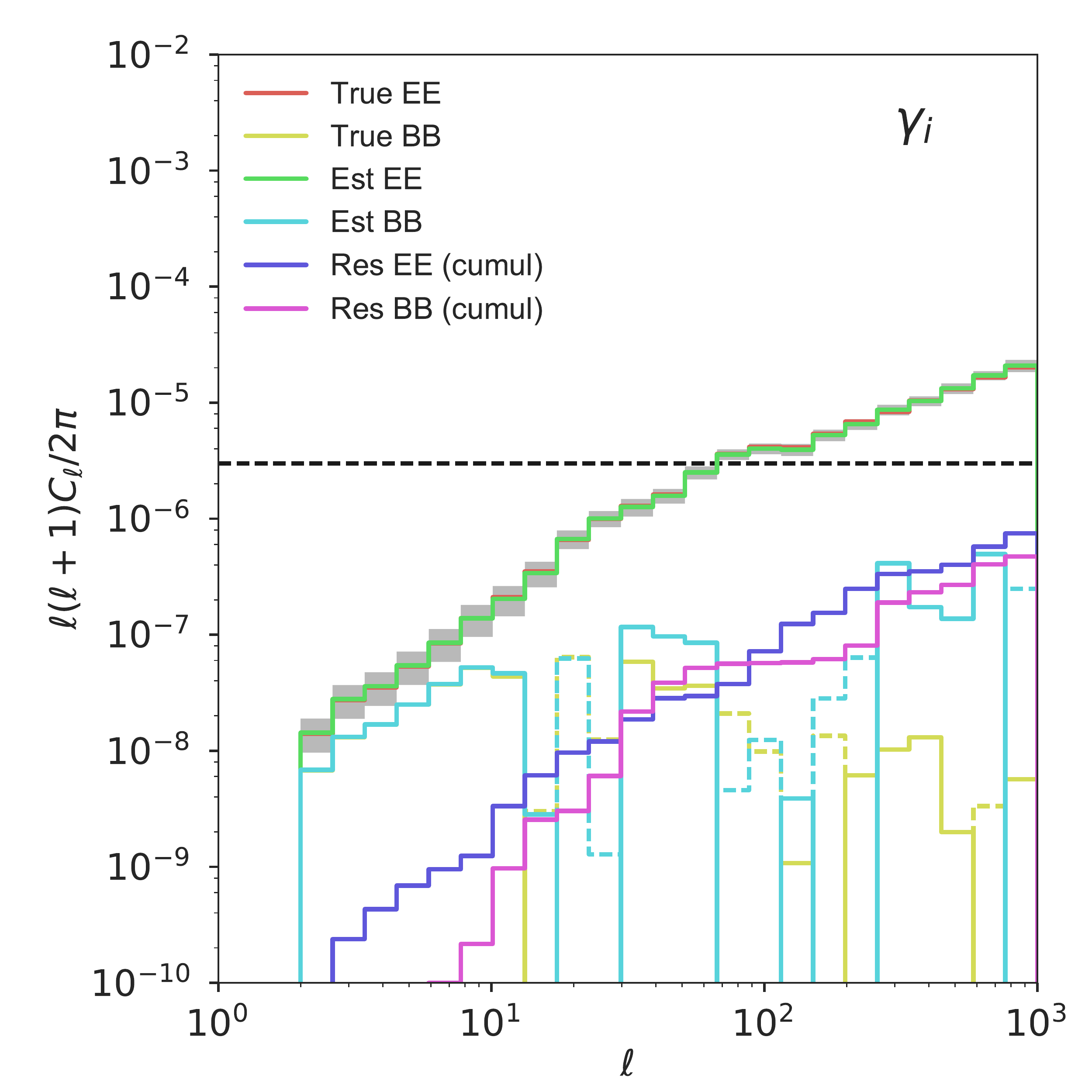}
  \caption{\label{fig:sheartwopt} Recovery of the true shear power spectrum $C_\ell^t$. The lines denote the E- and B-modes of the true shear map (True EE and True BB; red and yellow lines), the estimated E- and B-modes estimated by averaging 250 synthetic survey realizations (Est EE and Est BB; green and cyan lines), and the cumulative residual E- and B-modes (Res EE and Res BB; blue and magenta lines), which correspond to the $\sigma_{\rm sys}^2$-integral (equation~\ref{eq:sigmasyssqdef}). The red line denoting the E-modes of the true shear map are mostly under the green line (estimated E-modes). The dashed, black horizontal line shows the requirements target. The gray boxes around the estimated E-mode power spectrum shows the statistical uncertainty on the average of these 250 synthetic surveys.}
\end{figure}

\begin{figure}[ht]
  \centering
  \includegraphics[width=\textwidth]{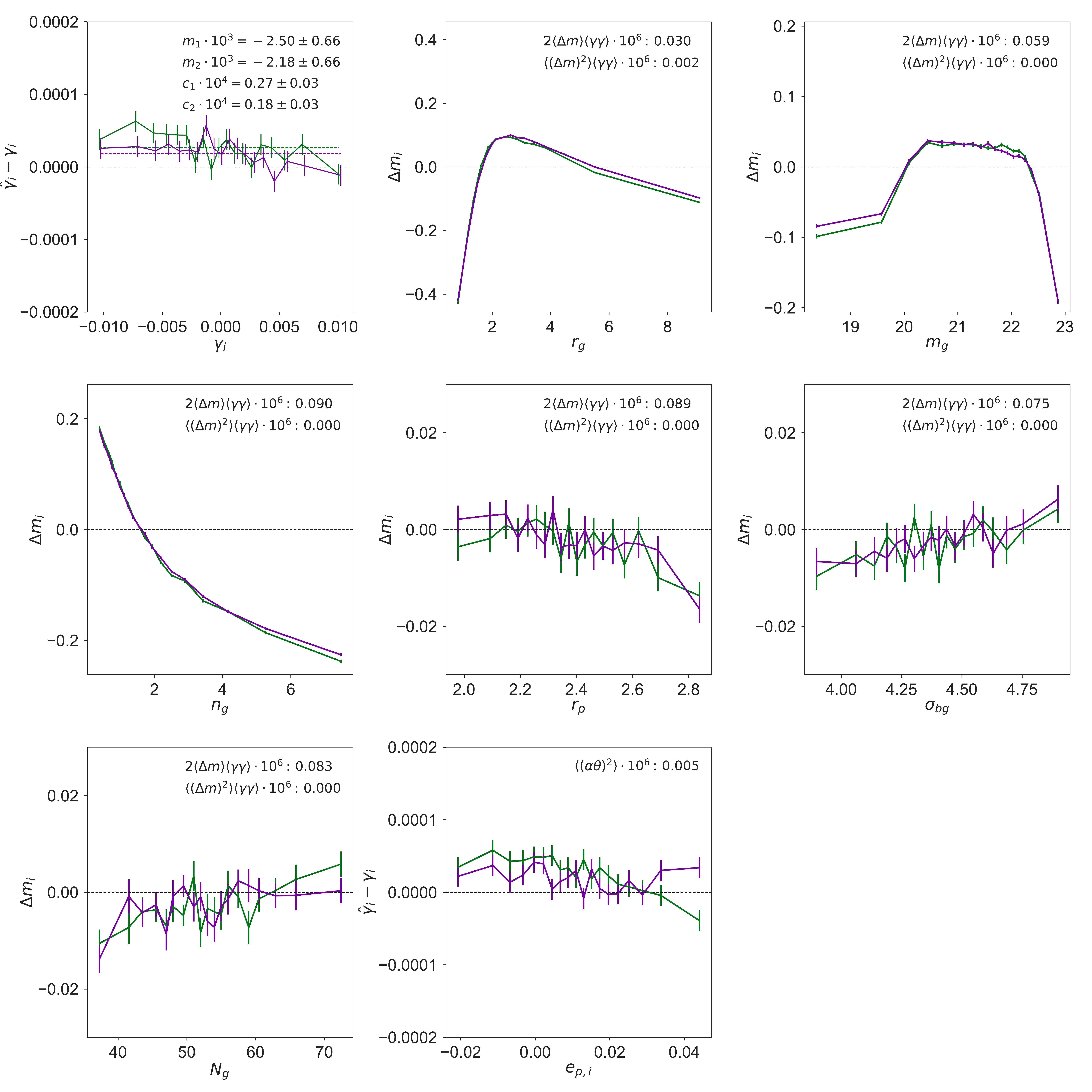}
  \caption{\label{fig:onept} One-point diagnostic functions from a combination of 250 synthetic survey realizations. The panels are from left to right, top to bottom: 1. Residual shear signal $\hat{\gamma_{i}}-\gamma_{i}$ vs. true shear $\gamma_{i}$ ($i \in {1, 2}$); 2.-7. Multiplicative biases in bins of a scalar intrinsic quantity (2. galaxy intrinsic half-light radius $r_g$; 3. galaxy intrinsic magnitude $m_g$; 4. S\`ersic index $n_g$; 5. PSF half-light radius $r_p$; 6. local rms of the background noise $\sigma_{bg}$; 7. local number of galaxies in 1.7$\times$1.7 arcmin$^2$ pixels $N_g$); and 8. residual shear signal vs. PSF ellipticity $e_{i, PSF}$ ($i \in {1, 2}$). The green and purple curves are for the 1- and 2-components respectively. The numbers in the upper right corner denote in the first panel the corresponding multiplicative and additive biases at the one-point level. In the other panels, they denote the contributions to $\sigma_{\rm sys,s}^2$, estimated on the diagnostic maps (1pt method; see section~\ref{sec:sigmasyseval}).}
\end{figure}

\begin{figure}[ht]
  \centering
  \includegraphics[width=\textwidth]{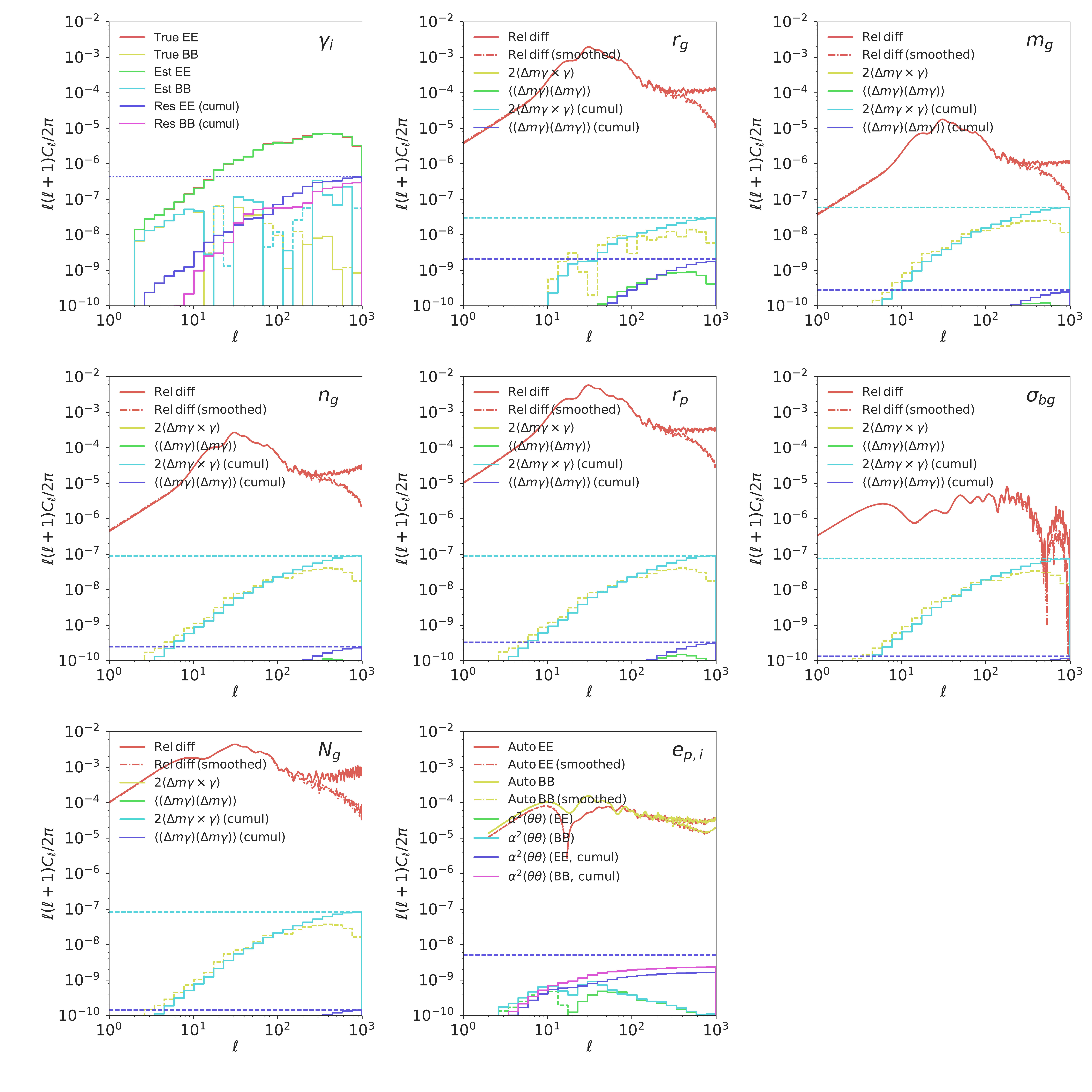}
  \caption{\label{fig:twopt} Two-point diagnostic functions from a combination of 250 synthetic survey realizations. The panels are the corresponding panels to figure~\ref{fig:onept}, but with two-point functions instead of one-point functions. The lines in the panels from left to right, top to bottom are: 1. E- and B-modes of the smoothed true shear power spectrum (True EE and True BB; red and yellow lines), E- and B-modes of the estimated shear power spectrum on smoothed maps (Est EE and Est BB; green and cyan lines), and the residual E- and B-modes (Res EE and Res BB; blue and magenta lines) (cf. figure~\ref{fig:sheartwopt}); 2.-7. Angular power spectrum of the $\Delta\theta/\bar{\theta}$ relative difference map before (red solid line) and after smoothing (red dashed line), the power spectra contributing to $\sigma_{\rm sys,s}^2$ (yellow and green lines; see section~\ref{sec:sigmasyseval}), and the cumulative of these power spectra (2pt method; cyan and blue lines); 8. E- and B-modes of the PSF auto-correlation before (red and yellow solid lines) and after smoothing (red and yellow dashed lines), and the power spectra (green and cyan lines) and their corresponding cumulative contributions to $\sigma_{\rm sys,s}^2$ (2pt method; blue and magenta lines). The horizontal, cyan and blue dashed lines in each of the panels 2.-8. show the 1pt-contributions to $\sigma_{\rm sys,s}^2$ for the corresponding systematics source (see figure~\ref{fig:onept}). The sum of all these values is the dashed blue line in the first panel.}
\end{figure}

As explained in section~\ref{sec:implementation}, an area larger than the target area of the measurement needs to be simulated in order to evaluate the systematic error contributions due to a miscalibration of the shape measurement. Thus, we have simulated 250 realizations of our synthetic survey each covering an area of about 1,700 deg$^2$, i.e. equivalent to about 10 full skies (for details see section~\ref{sec:syntheticsurveys} and table~\ref{tab:syntheticsurveytable}).

We apply the steps described in section~\ref{sec:shearmsrmtmethod} to this simulated data set. First, the star-galaxy separation is performed (section~\ref{sec:sgseparation}). Next, we select the galaxy sample and find a resulting galaxy density of about 4 per arcmin$^2$ (section~\ref{sec:galaxysample}). We then use a subset of 10 survey realizations to calibrate the shear measurement by comparing shear estimates to the input shear signal for each individual galaxy, yielding the calibration factors (section~\ref{sec:shearmsrmt})
\begin{equation}
  \lambda=0.5465, \, \, \, \, \, \, \eta=0.448475.
\end{equation}
We apply the calibration factors to the whole data set, and create maps for which we estimate the angular power spectra (section~\ref{sec:angularpowerspectrum}). Finally, we evaluate the systematic error contributions using the framework described in section~\ref{sec:implementation}.

The results are summarized in figures~\ref{fig:sheartwopt}-\ref{fig:twopt}. Figure~\ref{fig:sheartwopt} compares the true and the measured shear power spectra. The estimated E-mode power spectrum (green curve) lies on the true E-mode shear power spectrum $C_\ell^t$ (red curve). The power spectra differ at the level of about $\sim$1\%. The true (yellow curve) and estimated (cyan curve) B-mode power spectra are both significantly smaller. The cumulative contributions of these differences, interpreted as the contribution due to systematics, to $\sigma_{\rm sys}^2$ evaluated using equation~\ref{eq:sigmasyssqdef}, are shown by the blue and magenta lines. As described in section~\ref{sec:requirements}, we place a requirements target on $\sigma_{\rm sys}^2$. This target is given by the horizontal, black dashed line. The residual systematics in the recovery of the power spectrum satisfy this limit for both the E- and B-modes.

Figure~\ref{fig:onept} shows the one-point diagnostics. The top left panel shows the multiplicative and additive biases for each shear component (equation~\ref{eq:shearbias}) for the full lensing sample. The panels 2-7 from top to bottom, left to right, show the multiplicative biases as a function of the 6 scalar systematics sources: intrinsic half-light radius $r_g$, intrinsic magnitude $m_g$, S\`ersic index $n_g$, PSF half-light radius $r_p$, local value of the rms of the background noise $\sigma_{bg}$, and local number of galaxies in 1.7$\times$1.7 arcmin$^2$ pixels $N_g$. We use these curves to construct the systematics maps we use to predict their induced systematic errors. In the top right corner, we give the contributions to $\sigma_{\rm sys,s}^2$ estimated on the systematics maps (1pt). The last panel shows the residual shear signal as a function of the spin-2 systematics source, the PSF ellipticity $e_{i, PSF}$. We consider in this work only the contribution to the additive bias of this source, which we estimate by fitting a linear relation (equation~\ref{eq:spin2shearfit}). As we effectively calibrate the PSF deconvolution by minimizing the PSF leakage $\alpha$, the residual slope is small.

Using the sensitivities shown in figure~\ref{fig:onept}, we construct systematics maps and smooth following the prescriptions in section~\ref{sec:sigmasyscatalog}. On these maps, the contributions of each systematics source to the total residual systematics parametrized by $\sigma_{\rm sys,s}^2$ are computed by either directly evaluating the systematics maps (1pt), or computing their E-mode angular power spectra (2pt). Ideally, we expect our prediction of the systematics levels $\sigma_{\rm sys,s}^2$ to yield consistent results for both approaches. Furthermore, the sum of all contributions should agree with the total residual systematics estimated by comparing the smoothed estimated and true shear power spectra (equation~\ref{eq:clrecovery}).

Figure~\ref{fig:twopt} shows the estimated power spectra of the smoothed shear (top left) and systematics maps (other panels). The ordering of the panels corresponds to the one in figure~\ref{fig:onept}. The top left panel compares the shear E- and B-mode power spectra estimated on smoothed shear maps to the smoothed true spectra (cf. figure~\ref{fig:sheartwopt}). The differences are interpreted to be caused by systematics, and we show their cumulative contribution to $\sigma_{\rm sys,s}^2$. The final value is consistent with the predicted sum of the systematics for the E-modes (blue dashed line). For completeness, we also show the total residual B-modes power spectrum, which is similar to residual E-modes.

The other panels show the power spectra contributing to the total residual error. As in figure~\ref{fig:onept}, the first 6 panels are for the scalar and the last panel for the spin-2 systematics sources. We plot the angular power spectra of the unsmoothed and smoothed maps of these quantities. The systematics levels we predict for each source estimated with the two methods (1pt: dashed horizontal lines; 2pt; solid lines of the same colors) are furthermore shown. These methods agree for all of the systematics sources, with the exception of the spin-2 systematics source, the PSF ellipticity $e_{p,i}$. The latter contributions, as estimated by both approaches, are however at a significantly lower level than the other systematics sources. As mentioned above, they are suppressed by the empirical PSF correction. We note that the angular power spectra decay in power at large multipoles $\ell$ due to the maps being smoothed. Since the 1pt-computation integrates over all possible scales, the 2pt-computation however does not. Hence, smoothing the maps and thus suppressing small-scale power is a necessary condition for the two approaches to yield consistent results. The sum of all the cyan and blue dashed lines (1pt) in the systematics panels corresponds to the blue dashed line in the first panel.

Our predictions of the systematics yields consistent results for $\sigma_{\rm sys,s}^2$ for the two methods (1pt and 2pt). Furthermore, their sum agrees with the residual shear signal estimated on smoothed shear maps. We also require the overall effective residual shear systematics evaluated on unsmoothed maps $\sigma_{\rm sys}^2$ to satisfy the target (dashed black line in figure~\ref{fig:sheartwopt}). With these two conditions in combination we are able to satisfy the requirements described in section~\ref{sec:requirements} for our survey configuration.

\section{Conclusion}\label{sec:discussion}

We have presented an application of the MCCL framework to calibrate the shear measurement for two-point statistics. After deriving the requirements for a given weak lensing survey configuration, we described diagnostics based on one-point statistics that are used for this calibration process. We also showed how these diagnostics can be used to predict the contributions of different sources of systematics to the total shear power spectrum. 

We then applied this framework to perform an integration test in which an input shear power spectrum is propagated through synthetic image simulations in order to recover the shear power spectrum. We have generated 250 realizations of a synthetic survey, each covering 1,700 deg$^2$, which corresponds to about 425,000 deg$^2$ of imaging data. These images have then been analyzed, and the shear has been measured. This measurement has first been calibrated on a subset of these images ($\sim$ 17,000 deg$^2$), and then the performance of the shear measurement has been assessed on all the images. For this analysis we ignore PSF modeling errors and star contamination of the galaxy sample. With these simplifying assumptions, we find that the shear measurement can be calibrated within the required accuracy. Furthermore, we account for the total residual shear systematics in terms of the different contributing systematics sources.

This paper is part of a series on the implementation of the MCCL method \citep{Refregier:2014aa}. It extends the work of B16 from shear one-point to two-point functions. Another application is the measurement of the redshift distribution, which was given in \cite{Herbel:2017aa}. In future work, an application of the MCCL method to current surveys, including PSF modeling errors and other effects, will be presented.

\section{Acknowledgements}\label{sec:acknowledgements}

We thank Joel Akeret and Raphael Sgier for useful discussion. We acknowledge support by SNF grant 200021\_169130.

This project used public archival data from the Dark Energy Survey (DES). Funding for the DES Projects has been provided by the U.S. Department of Energy, the U.S. National Science Foundation, the Ministry of Science and Education of Spain, the Science and Technology Facilities Council of the United Kingdom, the Higher Education Funding Council for England, the National Center for Supercomputing Applications at the University of Illinois at Urbana-Champaign, the Kavli Institute of Cosmological Physics at the University of Chicago, the Center for Cosmology and Astro-Particle Physics at the Ohio State University, the Mitchell Institute for Fundamental Physics and Astronomy at Texas A\&M University, Financiadora de Estudos e Projetos, Funda\c{c}\~{a}o Carlos Chagas Filho de Amparo \`{a} Pesquisa do Estado do Rio de Janeiro, Conselho Nacional de Desenvolvimento Cient\'{i}fico e Tecnol\'{o}gico and the Minist\'{e}rio da Ci\^{e}ncia, Tecnologia e Inova\c{c}\~{a}o, the Deutsche Forschungsgemeinschaft and the Collaborating Institutions in the Dark Energy Survey. The Collaborating Institutions are Argonne National Laboratory, the University of California at Santa Cruz, the University of Cambridge, Centro de Investigaciones En\'{e}rgeticas, Medioambientales y Tecnol\'{o}gicas-Madrid, the University of Chicago, University College London, the DES-Brazil Consortium, the University of Edinburgh, the Eidgen\"{o}ssische Technische Hochschule (ETH) Z\"{u}rich, Fermi National Accelerator Laboratory, the University of Illinois at Urbana-Champaign, the Institut de Ci\`{e}ncies de l'Espai (IEEC/CSIC), the Institut de F\'{i}sica d'Altes Energies, Lawrence Berkeley National Laboratory, the Ludwig-Maximilians Universit\"{a}t M\"{u}nchen and the associated Excellence Cluster Universe, the University of Michigan, the National Optical Astronomy Observatory, the University of Nottingham, the Ohio State University, the University of Pennsylvania, the University of Portsmouth, SLAC National Accelerator Laboratory, Stanford University, the University of Sussex, and Texas A\&M University.

This research made use of \textsc{HOPE} \citep{Akeret:2015aa}, \textsc{IPython} \citep{Perez:2007aa}, \textsc{NumPy} and \textsc{SciPy} \citep{VanderWalt:2011aa}, \textsc{Matplotlib} \citep{Hunter:2007aa}, \textsc{Astropy} \citep{Astropy:2013aa}, \textsc{Seaborn}, \textsc{h5py}, and \textsc{healpy}.

\bibliography{Bibdesk}

\appendix

\section{Creating input PSF maps}\label{sec:creatingpsfmaps}
\begin{figure}[ht]
  \centering
  \includegraphics[width=.75\textwidth]{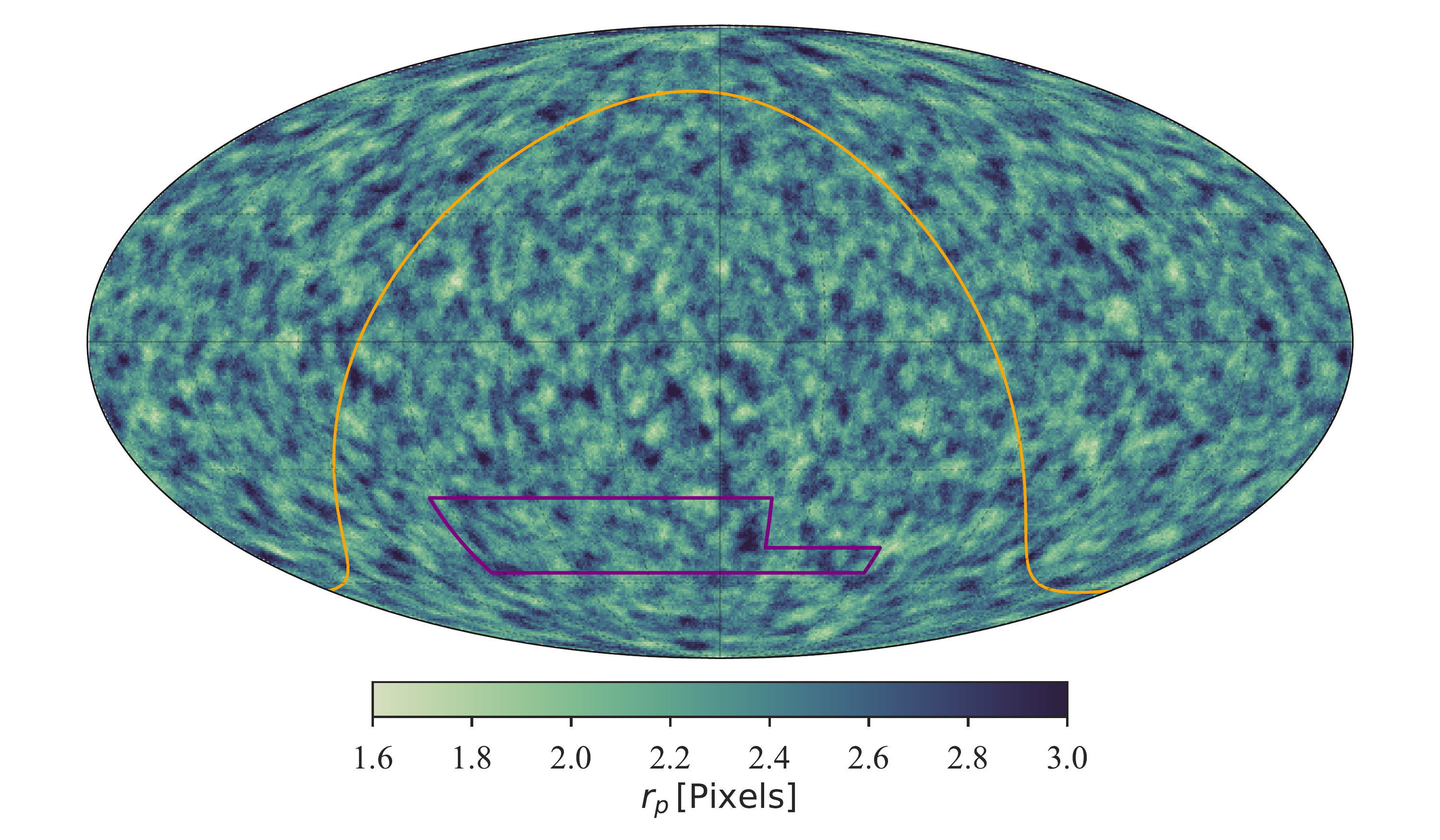}
  \includegraphics[width=.75\textwidth]{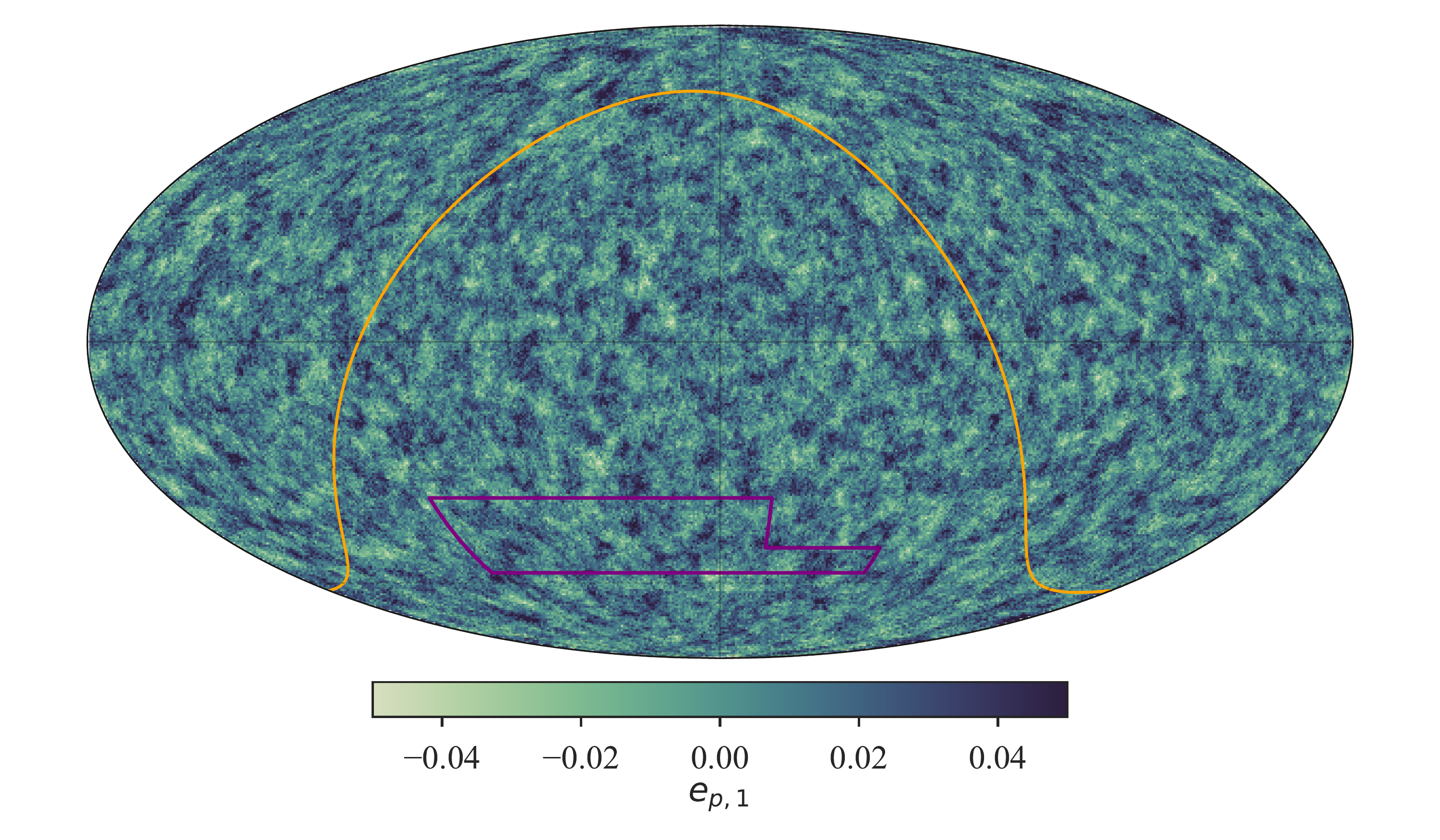}
  \includegraphics[width=.75\textwidth]{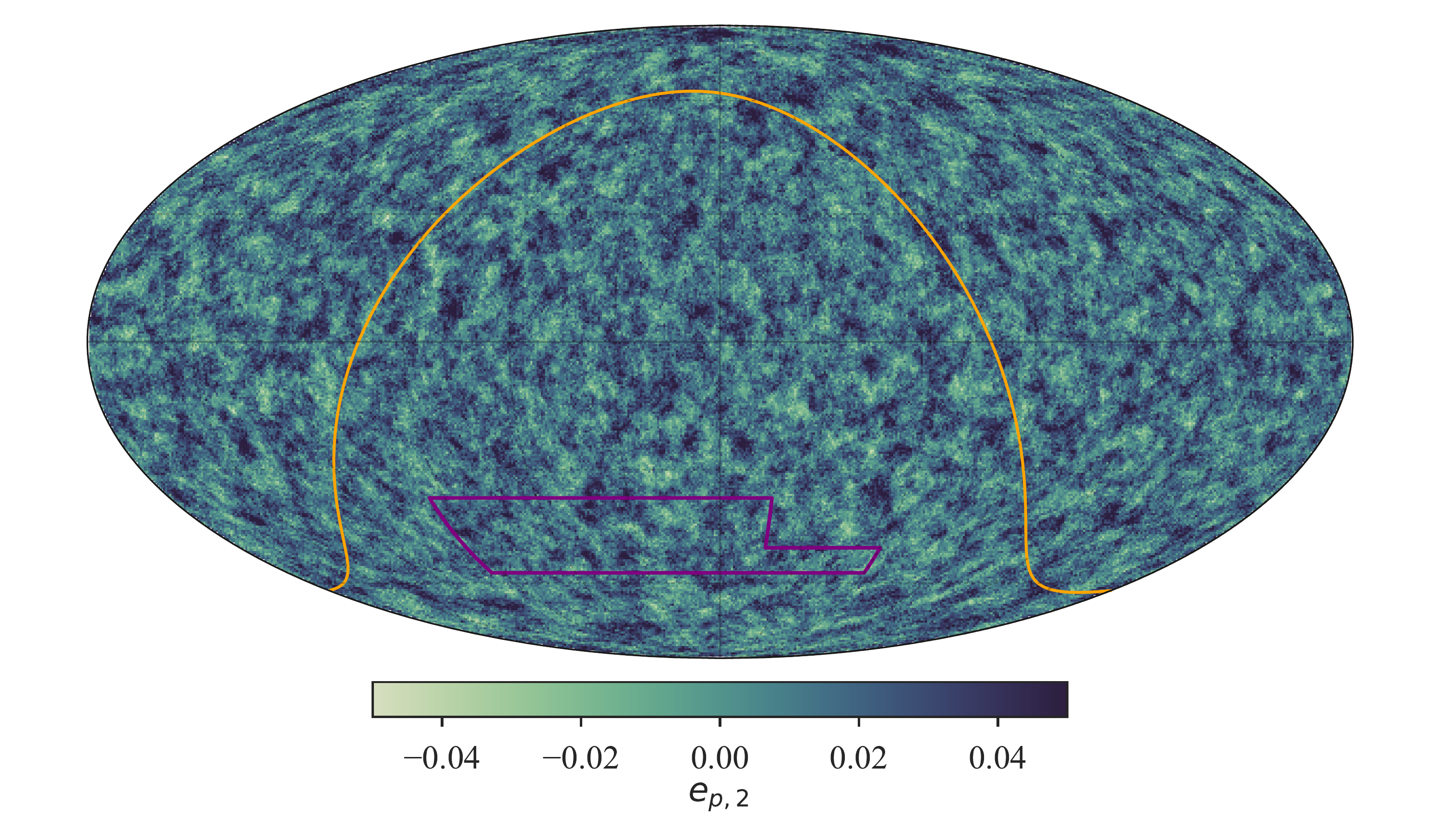}
  \caption{\label{fig:psfmaps} Generated PSF maps used as input to the simulations. The panels display from top to bottom: PSF half-light radius $r_p$; PSF ellipticity 1-component $e_{p,1}$; and PSF ellipticity 2-component $e_{p,2}$. In each of the panels, the orange line denotes the Galaxy, while the magenta region is the area our synthetic survey covers. The pixel scale for the first map is set to 0.263 arcsec.}
\end{figure}

We have used the \textsc{im3shape} catalog \citep{Jarvis:2016aa} on which we performed the same selection as detailed in section 9 of \cite{Jarvis:2016aa}. We then extracted the \textsc{MEAN\_PSF\_E1}, \textsc{MEAN\_PSF\_E2}, and \textsc{MEAN\_PSF\_FWHM} columns, which describe the mean PSF ellipticity and size across all the single exposures. Since the ellipticity definition employed in \cite{Jarvis:2016aa} corresponds to $(a-b)/(a+b)$, we have first converted the ellipticities to the ellipticity definition employed in this work.

Using these PSF estimates, we have created \textsc{HEALPix} maps with $\textsc{nside}=1024$ of the mean-subtracted PSF ellipticity, and a relative difference map of the PSF FWHM, which is dimensionless. On these maps we have then run \textsc{PolSpice}, setting $\textsc{thetamax}=15\, \mathrm{deg}$ and $\textsc{apodizesigma}=7.5$. Based on these power spectra, we have then adjusted the parameters in equation~\ref{eq:psfmodel} yielding the PSF model given in table~\ref{tab:syntheticsurveytable}. Finally, we have generated Gaussian random fields and then shifted and rescaled these maps by the mean PSF size and ellipticities (see section~\ref{sec:obsinsteffects}). Using a fixed Moffat parameter $\beta$ set to 2.5 we have furthermore converted the PSF FWHM map to a half-light radius map.

Figure~\ref{fig:psfmaps} shows the maps we have used in our analysis.

\end{document}